\documentclass{aa}
\usepackage{graphicx}
\usepackage{epsfig}
\usepackage{latexsym}
\usepackage{txfonts}
\begin{document}
\renewcommand{\labelitemi}{-}
\title{Relations between concurrent hard X-ray sources in solar flares}
\author{Marina Battaglia
  \and Arnold O. Benz}
\institute{Institute of Astronomy, ETH Zurich, 8092 Zurich, Switzerland}
\date{Received /Accepted}

\abstract
{Solar flares release a large fraction of their energy into
  non-thermal electrons, but it is not clear where and how. Bremsstrahlung
   X-rays are observed from the corona and chromosphere. }
{We aim to
  characterize the acceleration process by the coronal source and its leakage
  toward the footpoints in the chromosphere. The relations between the sources reflect 
the geometry and constrict the configuration of the flare.} 
{We studied solar
  flares of GOES class larger than M1 with three or more hard X-ray sources
  observed simultaneously in the course of the flare. The events were
  observed with the X-ray satellite RHESSI from February 2002 until July 2005.
  We used imaging spectroscopy methods to determine the spectral evolution of
  each source in each event. The images of all of the five events show two
  sources visible only at high energies (footpoints) and one source only
  visible at low energies (coronal or looptop source, in two cases situated
  over the limb). } 
{ We find soft-hard-soft behavior in both, coronal source and footpoints. The coronal
  source is nearly always softer than the footpoints. The footpoint spectra
  differ significantly only in one event out of five.}
{The observations are consistent with acceleration in the coronal source and an intricate connection between the corona and chromosphere.}

\titlerunning{Relations between X-ray sources in solar flares}
\authorrunning{Marina Battaglia \& Arnold O. Benz}

\maketitle


\section{Introduction} \label{Introduction}
The current understanding of solar flares leaves open fundamental questions such as: where is flare energy released, how are particles accelerated?
A large part of the energy released in a solar flare is converted into energetic electrons emitting hard X-rays. Therefore,
observations in X-ray wavelengths give quantitative measures of heating and
particle acceleration in the flare. X-ray observations by Hoyng et al. (\cite{Ho81})
showed hard X-ray (HXR) sources at both ends of a loop structure, commonly called footpoints. They are thick target bremsstrahlung
emission produced by precipitating electrons, accelerated somewhere
in the loop. Footpoints can also be seen in H$\alpha$ and EUV
(e.g. Gallagher et al. \cite{Ga00}, Fletcher et al. \cite{Fl04}), indicating
the precipitation of flare particles and the reaction of the thermal plasma. In an event observed by \textit{Yohkoh},
Masuda et al. (\cite{Ma94}) first noted a third HXR source 
situated above the looptop (looptop or coronal source). Alexander \& Metcalf (\cite{Al97}) analyzed this event carefully, concluding that the loop top source can be best described by a thermal component and a non-thermal component which is harder than the footpoint spectrum.
Petrosian et al. (\cite{Pe02}) made an extended study of looptop sources and footpoints in \textit{Yohkoh}-events. They find that the spectral index of the looptop source is softer than the footpoints on the average by about 1. The accuracy of their spectra however, was limited by the energy resolution of the \textit{Yohkoh} detectors. \\ 

An important observation about the time behavior of the HXR flux has 
already been made in the late 1960s by Parks \& Winckler
(\cite{Pa69}) and Kane \& Anderson (\cite{Ka70}).
They found that the hardness of a 
spectrum changes in time
and that there exists a correlation between the HXR flux and the hardness of
the spectrum (soft-hard-soft or SHS). These observations were later confirmed
by several authors, e.g. Benz (\cite{Be77}); Brown \& Loran (\cite{Br85}); Lin \&
Schwartz (\cite{Li86}); Fletcher \& Hudson (\cite{Fl02}); Hudson \& F\'arn\'ik
(\cite{Hud02}). Beside the SHS pattern, a soft-hard-harder (SHH) pattern has
also been observed in some events (Frost \& Dennis \cite{Fr71}; Cliver et al. \cite{Cl86}; Kiplinger \cite{Ki95}).
A quantitative study of 24 solar flares
observed by RHESSI on this subject has been made by Grigis \& Benz
(\cite{Gr04}). They find that elementary flare bursts also show SHS. Battaglia
et al. (\cite{Ba05}) made a study of flares of different size, finding that
events with smaller HXR flux are softer on the average and that the relation between
HXR flux and spectral index at peak time of events of different size is
the same as the one from several peaks of one event. 

Is the SHS-behavior a feature of the acceleration mechanism as previously claimed? Or is it a transport effect produced by collisions or return currents? A further possibility could be a change in the dominating
X-ray source from the coronal source (soft) to the footpoints (hard) and back
to the coronal source again (soft). Thus, is the SHS-behavior nothing but a coronal-footpoint-coronal effect? The previous studies have been made
using full sun spectra. To investigate the cause of the SHS, the spectra of each source must be analyzed separately.  
The Ramaty High Energy Solar Spectroscopic Imager (RHESSI, Lin et
al. \cite{Li02}) provides the possibility of making high resolution imaging spectroscopy at different locations on the
sun. One can therefore study each source separately in events with several
contemporaneous HXR-sources. The high energy resolution yields detailed spectra, allowing a reliable differentiation between thermal and non-thermal emission.
Emslie et al. (\cite{Em03}) made an analysis
of a very large event with 4 HXR-sources observed by RHESSI. 
They find a coronal
source with a strong thermal component and two (at times three) footpoints 
in regions with opposite magnetic
polarity.  They report that the spectral indices of the footpoints differ notably and accredit
this to collisional losses by different column densities in the loop connecting the footpoints to
the coronal source.

The purpose of this work is a systematic study of the relation between 
coronal source and footpoints
in time and spectra for several well observed events. The events were carefully selected, not necessarily the largest ones, but those with informative data concerning both, thermal
and non-thermal source parameters. The RHESSI data has been searched for well
separated, bright events without strong pileup, situated near the limb. We present here the results for the
best observed events of the first 40 months since launch.


\section{Observations, Event Selection and Spectroscopy} \label{eventselection}
The X-ray satellite RHESSI has been observing the full sun since February 2002. Modulation
of the X-ray flux by rotating grids provides image information for any region
on the sun (Hurford et al. \cite{Hu02}). High resolution germanium detectors
(energy resolution $\sim$ 1~keV) allow detailed studies of X-ray flare-spectra (Smith et
al. \cite{Sm02}). In Sect.~\ref{evselection} we describe how events were
selected. The image processing and spectral analysis methods are presented in
Sect.~\ref{imspec} along with some investigations of the best
choice of imaging algorithm, source regions etc. 

\subsection{Event selection} \label{evselection}

The selection was made using imaging spectroscopy quicklooks provided by the RHESSI 
Experimental Data Center (HEDC, Saint-Hilaire et al. \cite{Sa02}). 
Events are required to have 3 sources observed simultaneously during at least 
1 minute in the course of the flare. The sources may not all be visible in the same image of a particular energy range. The search was restricted to events
larger than GOES class M1 in order to have large enough count rates. Further, the three sources ought to be well separated to avoid contamination of spectra in imaging
spectroscopy by other sources. For this reason, we required a minimum offset of 700 arcsec from sun center to exclude events
with projection of the coronal source onto the footpoints. Events with strong 
particle precipitation and detector livetime (uncorrected monitor rates) below 90\% were
discarded. This lead to a final sample of 5 flares.
Table~\ref{evlist} gives an overview of the selected events.

\setlength{\tabcolsep}{1.2mm}
\begin{table}
\caption{List of analyzed events. The times give the range during which the
 analysis was made (times with strong emission from all three sources).}
\begin{center}
\begin{tabular}{lcr}   
\hline \hline
 Date & Time & GOES class \\
\hline
4-Dec-2002 & 22:42-22:53 & M2.7          \\
24-Oct-2003& 02:42-03:00 & M7.7       \\
1-Nov-2003 & 22:24-22:40 & M3.3             \\
13-Jul-2005 & 14:12-14:25 & M5.1       \\
30-Jul-2005 & 06:28-06:36 & X1.3          \\
\hline
\end{tabular}
\end{center}
 \label{evlist}
\end{table}

\begin{figure*}[!]
\centering
\includegraphics{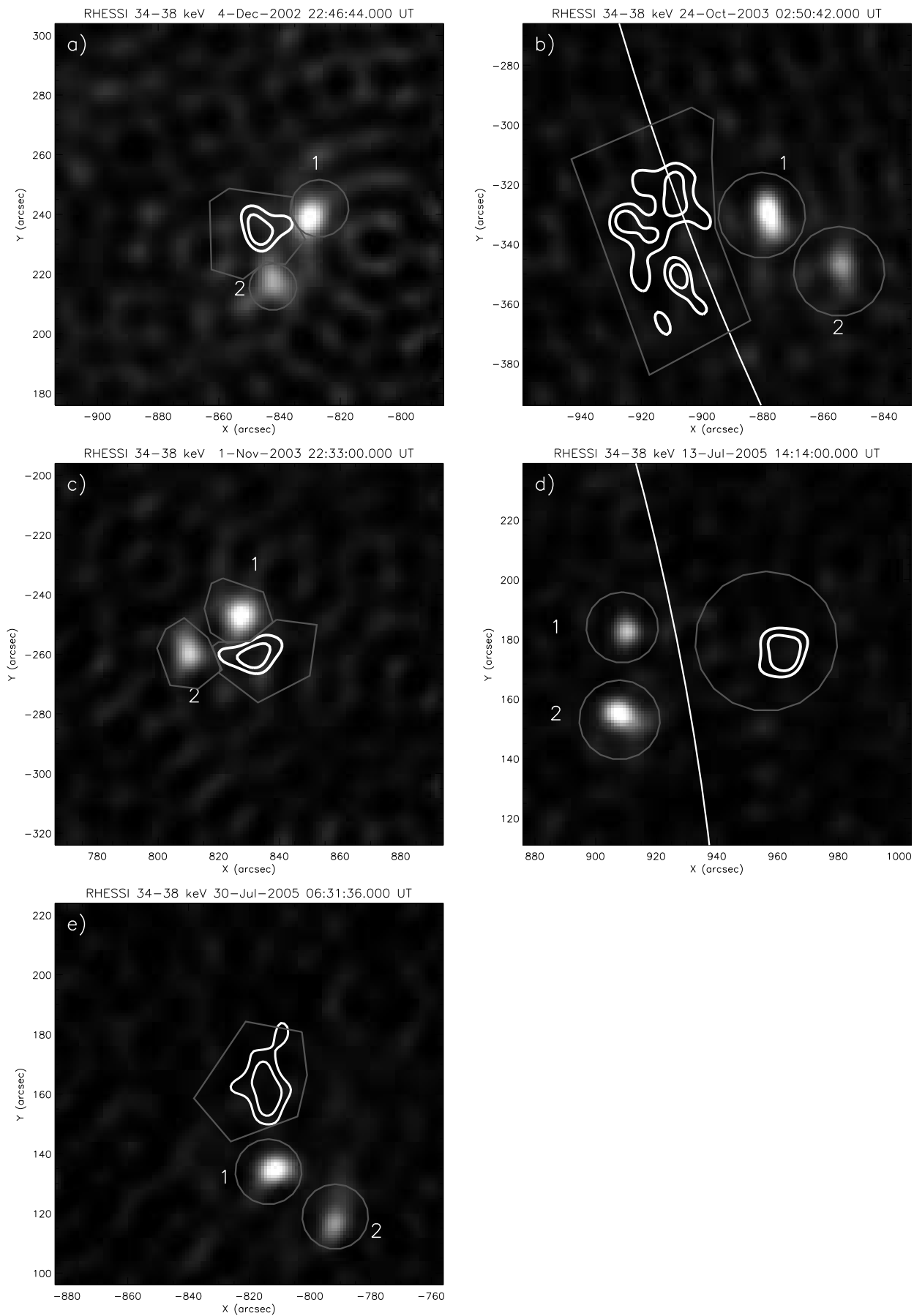}
\caption {Clean images (using detectors 3-8) of each event in the 34-38~keV energy range. The 60 and 80 \%
contours at energies 10-12~keV (white) and the regions of interest (grey) are given. The footpoints have
been arbitrarily numbered (1 \& 2). The solar limb is indicated where in the field of view.}
\label{events}
\end{figure*}

Grey-scale images of the events at 34-38~keV (representative for emission by non-thermal electrons) are presented in
 Fig.~\ref{events}. The  60 and 80 \% contours at
energies 10-12~keV (dominated by thermal emission) are overplotted. The regions of interest for the spectrum calculation are given in grey.
The fragmented shape of the coronal source of the Oct. 24th 2003 event can be
partly accounted for by over-resolution, as the source is slightly more compact
in images without detector 3. 

\subsection{Imaging spectroscopy} \label{imspec}
In this section we discuss some technical aspects of the analysis concerning
imaging spectroscopy as well as some issues that have to be considered like
source separation and pileup. 
\subsubsection{Imaging algorithm}
Clean, Forward-Fitting and
Pixon algorithms (Hurford et al. 2002) have been tested for image reconstruction. Clean was used for the actual imaging spectroscopy for the following reasons. Forward-Fitting works fine as long
as the sources in an image are equally strong, but has difficulties as soon as there are
background regions that are almost as strong as a source. With the defined 
time and energy bins for imaging spectroscopy, this frequently occurs in 
high or the lowest energy bands, at any time interval, in which case Forward-Fitting produces spurious results. 
Pixon yields generally a better spatial separation
of the sources than Clean. However, it needs more fine-tuning of the input imaging parameters to be as stable as Clean for low signal to noise ratios. 
In an extended series of tests, Clean turned out to be the most efficient and reliable algorithm for the automatic image generation of long time series for imaging spectroscopy.

Therefore, the Clean algorithm was applied, using detectors 3-8. Detector 2 over-resolves the sources, just increasing noise. The angular resolution without detector 3 
becomes too small to separate the sources properly for all events except the one
of July 13th 2005.  
The time bins were chosen from 12 s to 120 s, depending on the source 
intensities, to get good images and enough counts for reliable spectra. A pseudo-logarithmic energy binning was used. 

\subsubsection{Computation of spectra}\label{speccal}
\begin{figure*} 
\centering
\includegraphics{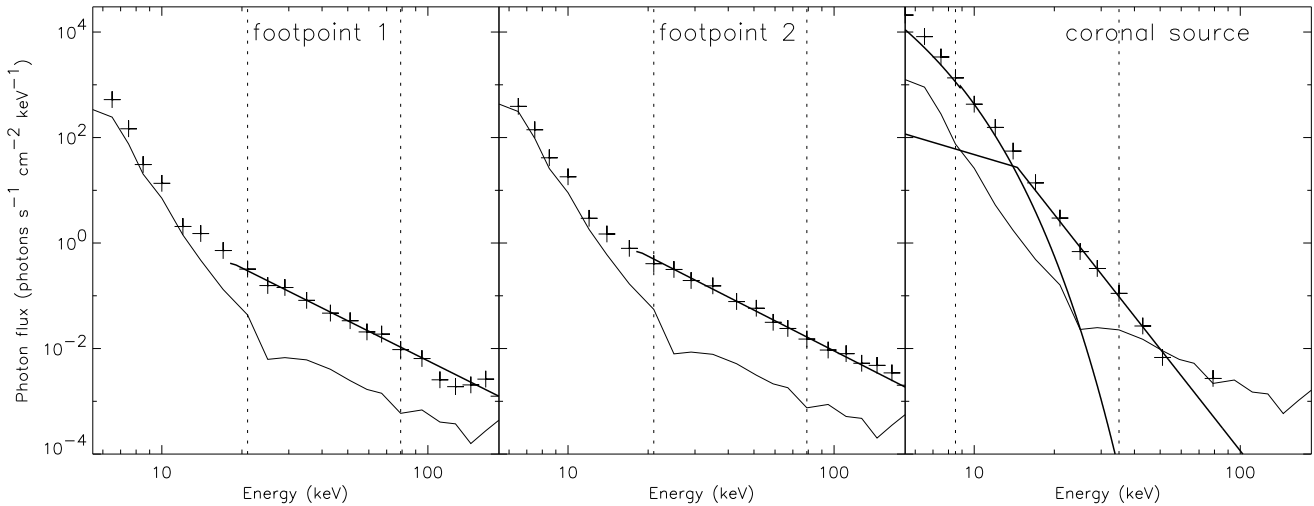}
\caption {Spectra of footpoints and coronal source for the time interval
  14:16:48-14:17:36 of the event of July 13th 2005. The energy range for the spectral fitting is indicated by the dotted lines. The thin solid line gives
  an estimate for the noise level (see Sect.~\ref{speccal}).}
\label{samplespec}
\end{figure*}

\begin{figure}
\resizebox{\hsize}{!}{\includegraphics{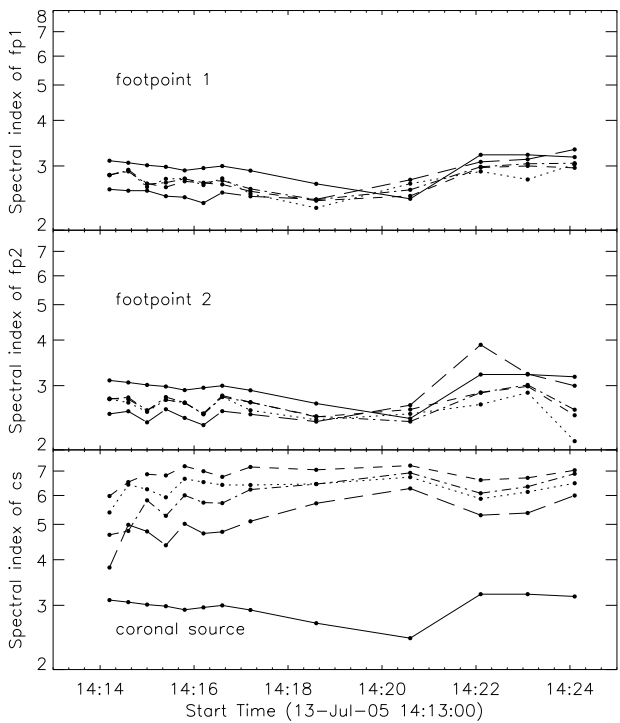}}
\caption {Time evolution of spectral index $\gamma$ of footpoints (\textit{top} and \textit{middle}) 
and coronal
  source (\textit{bottom}) for
  different choices of detectors and regions of interest. \textit{Solid}: full sun; 
  \textit{dotted}: with detector 3; \textit{dot-dashed}: without detector 3, same ROIs
  as dotted; \textit{dashed}: without detector 3, different ROIs; \textit{long dashes}: images with natural weighting of detectors.}
\label{test}
\end{figure}

The imaging spectroscopy tools implemented in 
OSPEX (a further development of SPEX (Schwartz \cite{Sc96}, Smith et al. \cite{Sm02})) have been used for the determination of the regions of interest (ROI) and for 
the calculation of the spectra. ROIs can be delimited by polygons or circles around the sources, as selected by the user. 
Ideally, one would select a ROI as contour in percentage of the maximum of a source. As sources move in time and often have different sizes in different energy-bands, this could not be done reasonably. Therefore, ROIs were selected as circles or polygons in each time interval, having the same size in all energies. The ROIs for the footpoints were defined at energies larger than 40~keV such as to include all of the source emission. The ROIs of the coronal source was selected at an energy around 10~keV likewise. 
The effects of this method of ROI selection are discussed in the last paragraph of this section and shown in Fig.~\ref{test}.

The attained spectra were fitted with a non-thermal power law at high
energies and a thermal component at low energies, where this was
possible. Some of the footpoints did not have any flare emission at lower
energies i.e. no measurable thermal emission. In this case only a power law was fitted to the energy range in which the flare emission was stronger than the noise level. 

Figure~\ref{samplespec} shows the spectra of the footpoints (fp 1 \& 2, left
and middle) and the coronal
 source (cs, right) for the time interval between 14:16:48-14:17:36 of the
July 13th 2005 event. The fitted power-law and thermal components are also
presented in Fig. \ref{samplespec}. 
For each time interval and each source, we made an estimate of the noise level in the Clean images. A spectrum from a large part in the image that had not been assigned to a source has been calculated and normalized for the area of the individual source ROIs. The result is indicated by the thin lines in
Fig.~\ref{samplespec}. This method will probably overestimate the actual noise, but can be used as a rough guide for the determination of the trustworthy energy intervals for the fitting (indicated by the dashed lines in Fig.~\ref{samplespec}).

The influences of the source delimitation and detector selection have been studied extensively. 
For the event of July 13th 2005, tests have been made with
differently defined ROIs, and with images with and without detector 3. Finally, a series of images with natural instead of uniform detector-weighting has been compared. Natural weighting gives the same weight to each subcollimator, opposed to uniform weighting where the collimators are weighted inversely proportional to their resolution, therefore giving the finer grids more weight.
Figure~\ref{test}
shows the time evolution of the spectral index $\gamma$, fitted to spectra
calculated for different choices of regions, detectors and weighting. The time
evolution of the spectral index of full-sun spectra in the same time bins has
been given for comparison. Four different cases were studied. 
Images with
detector 3, different ROIs around the same sources, and images without
detector three using the same ROIs as in the case with detector 3, as well as images with natural detector weighting. From the
time evolution of $\gamma$ one can see that the differences are small for the
footpoints. The quantitative differences for the coronal source are somewhat larger. They may be used
for an estimate on the error range of the spectral fittings. Note further
that the non-thermal component of the full-sun spectrum is mostly
due to footpoint emission. The coronal source causes a small shift
toward softer spectral indices, as expected. 

The qualitative behavior of the time evolution and the conclusions drawn from it do not change for the different approaches.\\

\subsubsection{Source separation}
The major problem encountered in the event selection was the sufficient separation of the sources. Many nice events had to be discarded because footpoints and
coronal source overlap. The
separation can be improved to a certain extent by optimizing parameters in the image
reconstruction,
 but in the end one is limited by the flare morphology. Usually the
footpoints are distinct and well defined.
 If any of the ROI's defined as described in section \ref{speccal}
overlapped, the event was not selected. In the event of Nov. 1st 2003 (Fig. \ref{events}c)
the ROIs come close, as the coronal source is embedded in a loop, visible at
10 keV, that extends nearly all the way to the footpoints. In this situation,
the spectrum of the non-thermal component of the coronal source may be
influenced by the emission of the footpoints. A similar case is the event of
Dec. 4th 2002 (Fig. \ref{events}a). The source separation in the other three events seems
clearly big enough to exclude an influence on each other.

\setlength{\tabcolsep}{1.0mm}
\begin{table*}[!]
\caption{Mean difference in spectral index $\gamma$ between footpoints and
  between coronal source and footpoints for all events in which it could be determined. 
Pivot energy $\mathrm{E^{piv}}$ for all
  sources and all events (where determinable). Temperatures derived from full sun spectra.}
\begin{center}
\begin{tabular}{llll|rrr|r}   
\hline \hline
 Date & $\mathrm{\gamma_{fp1}-\gamma_{fp2}}$ & $\mathrm{\gamma_{cs}-\gamma_{fp1}}$ &
 $\mathrm{\gamma_{cs}-\gamma_{fp2}}$ & $\mathrm{E^{piv}_{fp1}}$ & $\mathrm{E^{piv}_{fp2}}$&
 $\mathrm{E^{piv}_{cs}}$ & Temperature (MK)\\
\hline
4-Dec-2002 & -0.53 $\pm$ 0.20 & 1.22 $\pm$ 0.20 & 0.68 $\pm$ 0.15 & 13.74$\pm$0.41 &
 14.97$\pm$0.63 & 18.12$\pm$0.25 & 18.30 $\pm$ 1.68 \\
24-Oct-2003 & 0.33 $\pm$ 0.04 & 2.43 $\pm$ 0.22 & 3.07 $\pm$ 0.27 & - & - &
 22.74$\pm$2.99 & 22.51 $\pm$ 0.05 \\
1-Nov-2003 & -0.095 $\pm$ 0.093 & 0.72 $\pm$ 0.16 & 0.59 $\pm$ 0.24 & 14.68$\pm$1.14 &
 14.00$\pm$1.33 & 15.90$\pm$2.36 & 20.65 $\pm$ 1.99      \\
13-Jul-2005 & 0.13 $\pm$ 0.07 & 3.55 $\pm$ 0.13 & 3.68 $\pm$ 0.14 & - & - & -
 & 24.95 $\pm$ 1.07\\
30-Jul-2005 & 0.13 $\pm$ 0.07 & 1.15 $\pm$ 0.38 & 1.12 $\pm$ 0.41 & - & - &
 23.92$\pm$2.37 & 24.85 $\pm$ 0.31 \\

\hline
\end{tabular}
\end{center}
 \label{differences}
\end{table*}

\subsubsection{Pileup}\label{pileup}
Although the flares in our sample are not the largest ones and events with detector livetime (uncorrected monitor rates) below 90\% had been discarded in the selection, one still has to
consider the possibility of pileup (Smith et al. \cite{Sm02}) in certain time intervals. 
It does not play a substantial role in the footpoints as they
are observed and fitted above the energies where pileup is
worst. However, the non-thermal
part of the coronal source is observed at energies where pileup might cause
problems. We tested the importance of pileup in our events, using the
\texttt{hsi\_pileup\_check} routine. This routine calculates the corrected
  (counter) livetime, the effective pileup counts and the relation between corrected and uncorrected count spectra. Further, we examined images for a
"ghost"-source at the position of the coronal source at higher energies. 
Further, we compared the time
evolution of pileup flux to the time evolution of the coronal source flux at
the same energy (25~keV for attenuator state 1). The event of July 30th 2005
has attenuator state 3 throughout the observed time interval and shows no sign
of significant pileup. In some of the other events (all attenuator state 1),
pileup is a concern. For some times during the event of Nov. 1st and the end
of Oct. 24th 2003, more than
about 50 \% of the observed coronal HXR emission in the range between 20 and 30~keV has to be accounted
for by pileup. These times were not used in the further analysis (missing data in Fig.~\ref{SHS}).


\section{Results}

\subsection{Soft-Hard-Soft (SHS)} 
First we present the study of the SHS-behavior of individual sources.

\begin{figure*}[!]
\begin{minipage}[l]{0.5\textwidth}
\includegraphics[width=75mm]{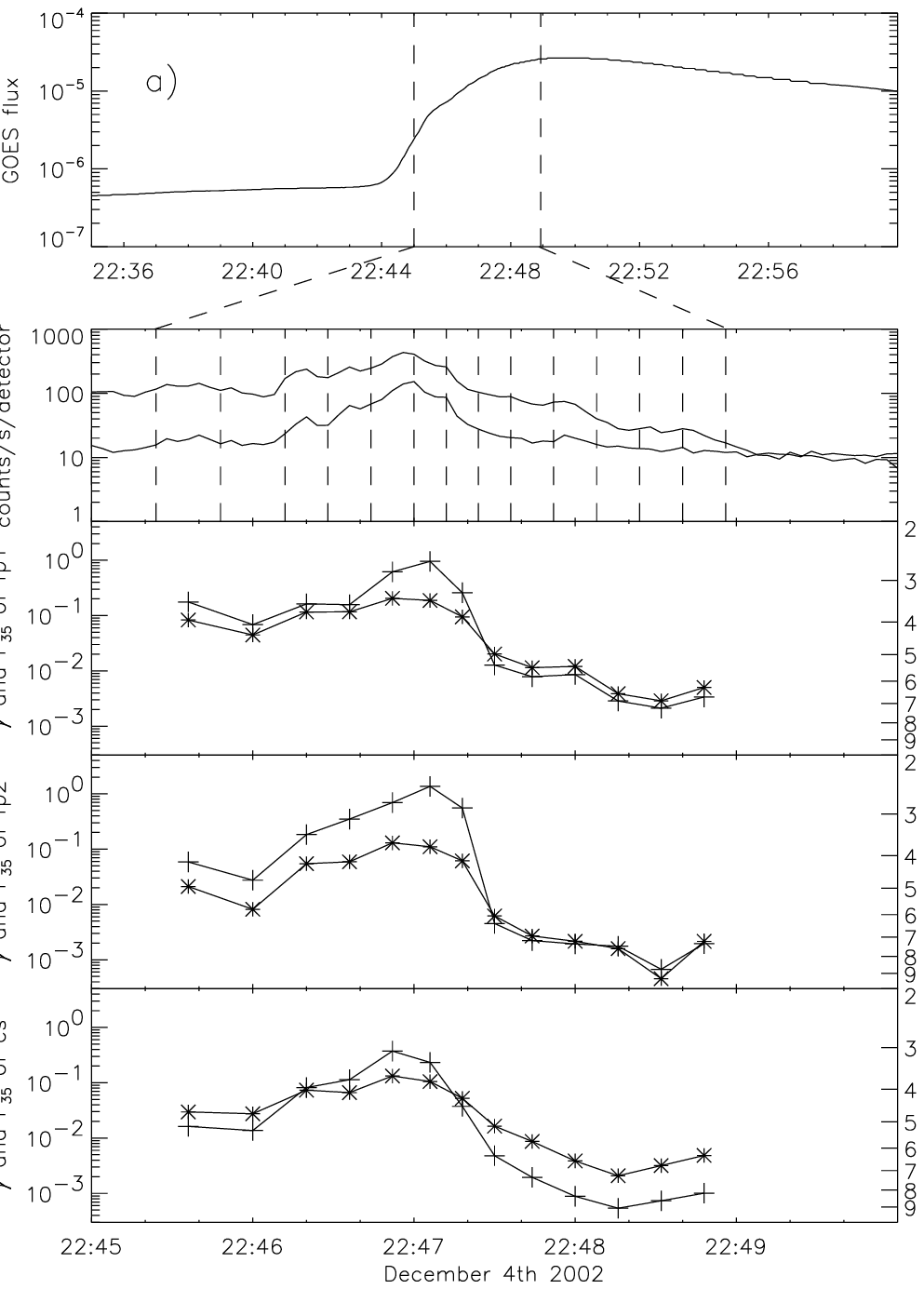}
\end{minipage}
\begin{minipage}[r]{0.5\textwidth}
\includegraphics[width=75mm]{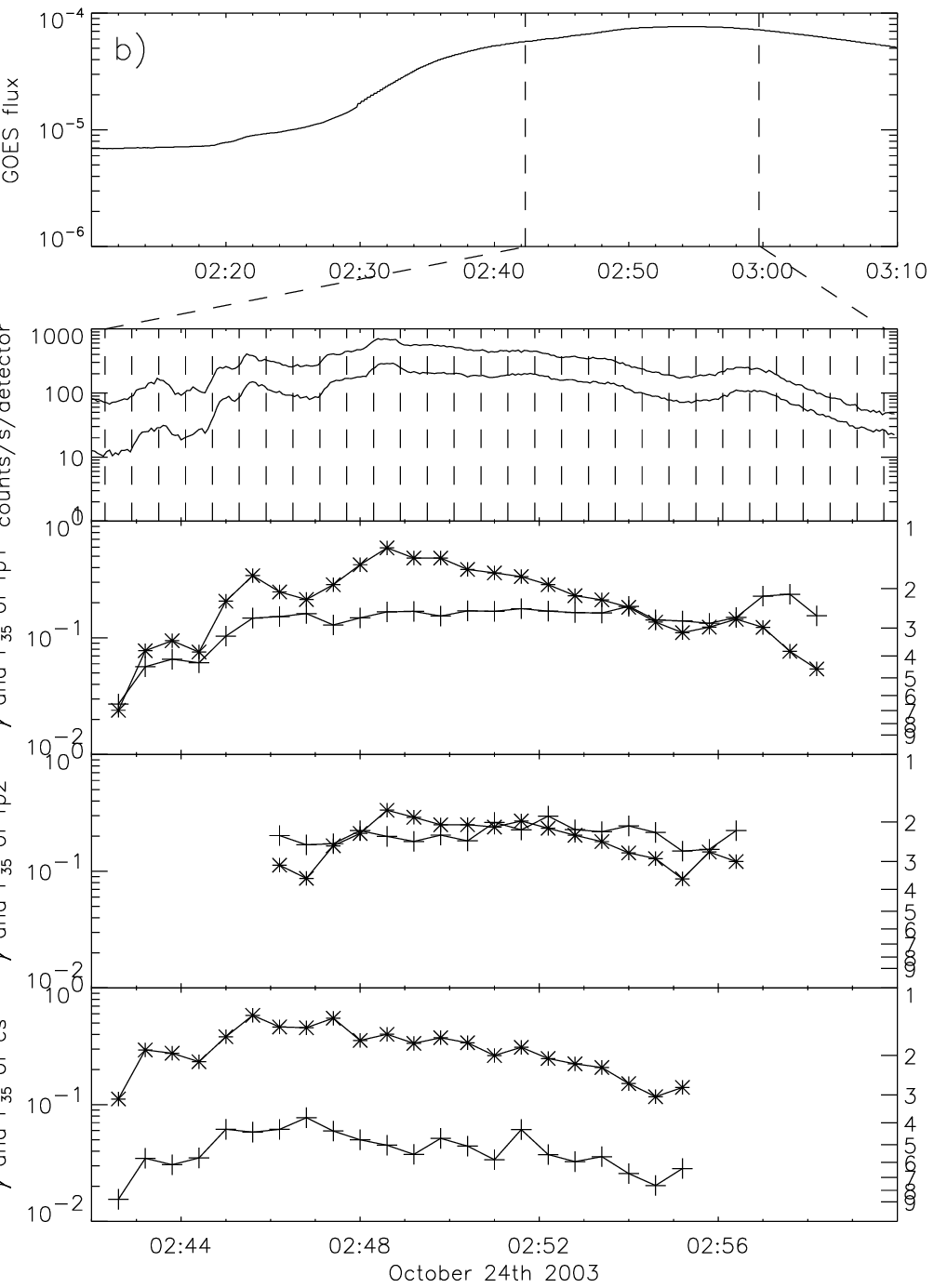}
\end{minipage}
\begin{minipage}[l]{0.5\textwidth}
\includegraphics[width=75mm]{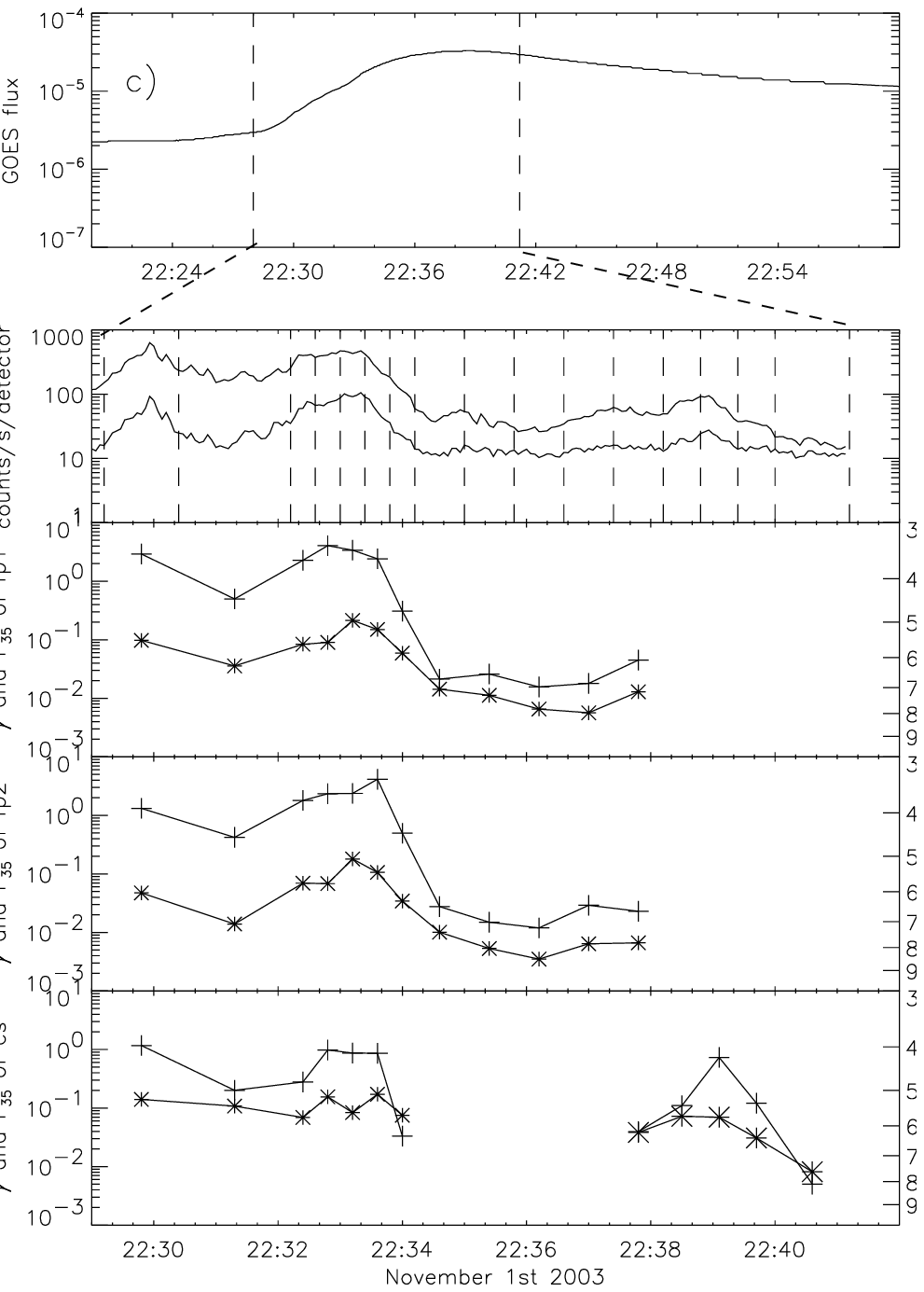}
\end{minipage}
\begin{minipage}[r]{0.5\textwidth}
\includegraphics[width=75mm]{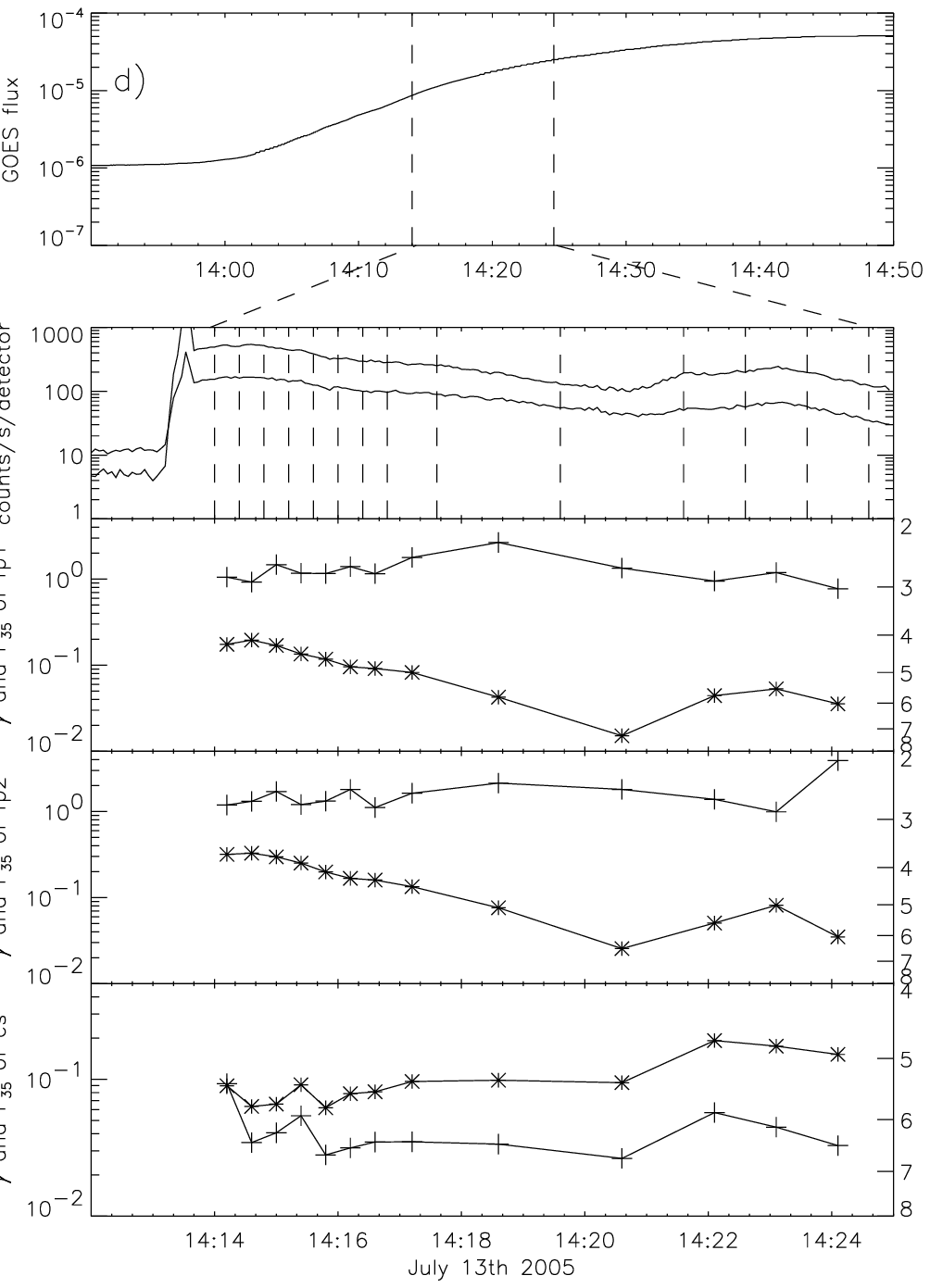}
\end{minipage}
\caption{\textit{Top}: GOES low-channel lightcurve. The dashed lines mark the analyzed time intervals. \textit{Lower four panels}: RHESSI full sun
lightcurves in energy-bands 25-50~keV and 50-100~keV, overplotted
is the time binning used for imaging spectroscopy (dashed lines); time evolution of spectral index ($\gamma$, \textit{crosses}) and non-thermal flux at 35~keV (F$_{35}$ [photons $s^{-1} cm^{-2} keV^{-1}$], \textit{stars}) for footpoints 1\& 2 and coronal source (bottom panel). The times
between 22:34:30 and 22:37 of the Nov. 1st 2003 event had to be neglected for the
coronal source because of pileup.}
\label{SHS}
\end{figure*}

For 4 events, the RHESSI full-sun count lightcurves in the energy-bands 3-12~keV, 25-50~keV and
50-100~keV are shown in Fig~\ref{SHS}. The time evolution of the fitted
non-thermal flux at 35 keV
(F$_{35}$) and the spectral index $\gamma$ are plotted for each
source. The variation in flux of the July 30th 2005 event (not shown in Fig.~\ref{SHS}). F$_{35}$ of the footpoints correlates well with the total count flux in the 25-50 and 50-100~keV energy bands, indicating that the spectral fits are plausible. 

In all previous observations of SHS-behavior (see Introduction) the full sun
spectrum has been analyzed. As previous measurements were made at relatively
high energies to avoid a contribution of the thermal component, they
predominantly refer to the footpoints. For the first time, it has become
possible to study the temporal evolution of the non-thermal component of the coronal source. 

In all events the coronal source varies clearly according to SHS (Fig.~\ref{SHS}). 
Three out of five events also show a more or less pronounced SHS-behavior in the footpoints, although there are times when the pattern is not very clear, or flux and spectral hardness anti-correlate. The events of Nov. 1st 2003 and Dec. 4th 2002 show a clear SHS-behavior in all sources.
 For Oct. 24th 2003, the variation in the flux is small without strong
peaks. The event of July 13th 2005 is peculiar. There is an
anti-correlation between flux (both, total count flux as well as fitted flux)
and spectral hardness in the footpoints. 

\subsubsection{Pivot point}
\begin{figure}
\resizebox{\hsize}{!}{\includegraphics{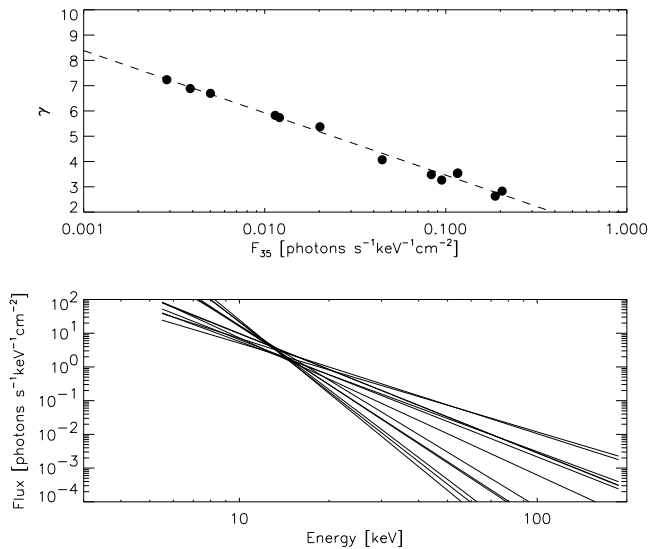}}
\caption {Example for the determination of the pivot point (for footpoint no.1 of Dec. 4th 2002). {\it Top:} flux and spectral index of all time intervals. The locations for the flux and spectral index in the best fitting pivot point model is shown {\it dashed}. {\it Bottom:} non-thermal power-law components for
all time intervals overplotted. The crossing point agrees with the pivot
point found by the above fit.}
\label{pivot}
\end{figure}

The first notion of an invariable point in solar flare spectra was made by Gan (\cite{Ga98}). 
This point was termed pivot point and analyzed quantitatively for the first time by  Grigis \& Benz (\cite{Gr04}, \cite{Gr05}). The SHS-behavior, indicating spectral hardening at large
fluxes, suggests that the non-thermal spectra at different times intersect at
a fixed point in energy and flux. Grigis \& Benz (\cite{Gr04}) noted that the
intersections of all spectra in an event are within a relatively small range of energies. Its average was termed pivot energy. We applied
the fitting method they describe in (\cite{Gr05}) to determine the pivot energy for each source. An example is shown in Fig.~\ref{pivot}. The results are given in Table~\ref{differences}. The physical significance of the pivot energy is not clear. However, it may be useful to describe the SHS-behavior quantitatively .

A pivot point could not be found for every source. 
If the variation in the flux and spectral index is
small, the power-law lines are nearly parallel in log-log, and the pivot point
is not well defined or does not exist. For one of the two cases
where all pivot points could be determined, the pivot energy of the coronal
source is higher by 3-5~keV than the pivot energies of the footpoints (see Table 2). In the other case, the three pivot energies are equal within errors. In the two cases where only the pivot energies of the coronal sources were found, the values even exceed 20~keV. 
All pivot energies for both, coronal source and footpoints given in Table 2 are higher than the mean value of 9~keV found by Grigis \& Benz (\cite{Gr04}) for
full sun spectra. The main contribution of non-thermal emission in full sun spectra originates from the footpoints. The pivot energies of the footpoints reported in Table 2 are outside the range of the half-power distribution of 6.5-12.5~keV reported by Grigis \& Benz. However, the deviation is statistically not significant.

\subsection{Difference in spectra between sources of the same event }
Non-thermal spectra in the 20-50 keV range can usually be well approximated by
a power law (Fig. \ref{samplespec}). The flux at a given energy and the spectral index thus
characterize the spectrum of a source. In this subsection the differences between spectral indices of the coronal
source and the footpoints, and between the two footpoints of each event are investigated . 
The mean differences in $\gamma$ time-averaged over the event are given in Table~\ref{differences}.

\subsubsection{Relation of coronal $\gamma$ to footpoint $\gamma$}
The coronal source is softer than both footpoints in all events at nearly all times (Fig. 4).  The smallest mean  difference of 0.59$\pm$0.24 was found for the event of Nov. 1st 2003 for which there is a possibility of source overlap. The maximum mean
difference is 3.68$\pm$0.14 for the event of July 13th 2005. Table~\ref{differences} points out that $\mathrm{\gamma_{cs} - \gamma_{fp}}$ in the 3 well separated events is
remarkably larger than in the two more compact events. Note in particular,
that the difference between coronal source and footpoints often differs
significantly from 2, the value expected from the difference between thick and
thin target sources. Nevertheless, the average over all mean differences is
1.82 with a standard deviation of 1.52. The weighted average and mean error are
1.98 $\pm$ 0.42. This finding does not agree with previous reports (Petrosian
et al.~\cite{Pe02}) based on \textit{Yohkoh} data. A possible explanation for the larger value is our selection of spatially separated sources, avoiding overlap between them. 

\subsubsection{Differences between footpoints}
Emslie et al. (\cite{Em03}) reported differences of 0.3-0.4 between the
spectral indices of the two stronger footpoints in their event. 
For the flares analyzed here, a significant difference is found in only one
out of five events, the Oct. 24th 2003 flare. For all other events, the mean difference in $\mathrm{\gamma_{fp}}$ is zero within the statistical uncertainty. 

\begin{figure*}
\resizebox{\hsize}{!}{\includegraphics{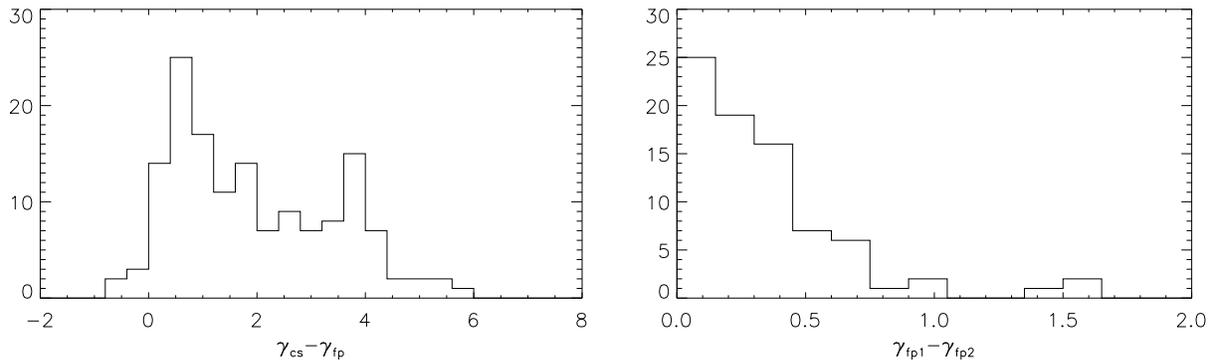}}
\caption {Distributions of the differences in spectral index ($\gamma$)
between coronal source and both footpoints of all events at all time bins
({\it left}) and between footpoints only ({\it right}). For the differences between footpoints, absolute values are shown.}
\label{diffdist}
\end{figure*}

Figure~\ref{diffdist} shows the distributions of the differences in the
spectral indices of the non-thermal emission as measured in all time bins and
all events. The difference between coronal source and footpoint spectral index
is almost always larger than zero (Fig.~\ref{diffdist}, left). The question of the transition from footpoints to coronal source will be addressed in section 3.3.
The differences between the footpoints are given in absolute values, as the footpoint numbering is arbitrary. 
As expected from the observations of the individual events (Fig. 4), the
distribution peaks at zero.

\begin{figure}
\resizebox{\hsize}{!}{\includegraphics{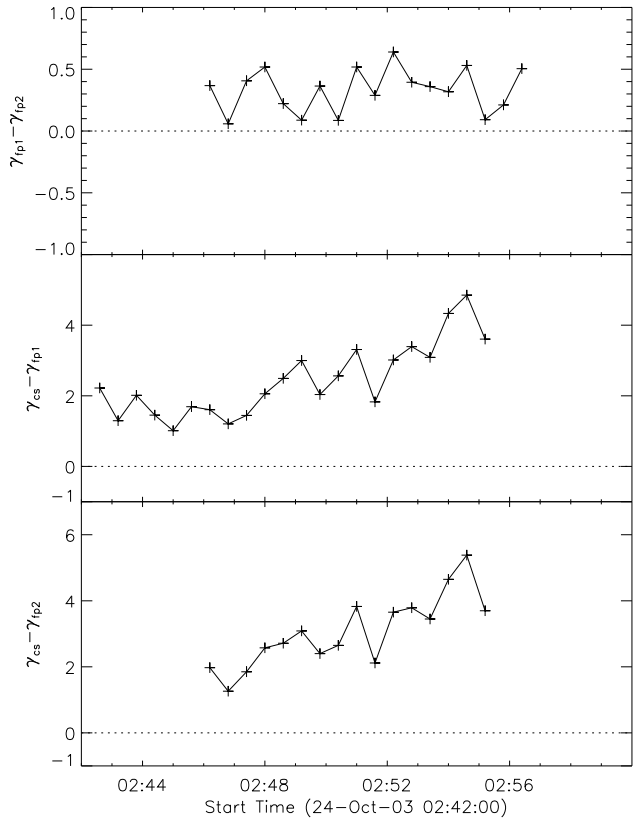}}
\caption {Time evolution of the difference between spectral indices of the three
 sources for the event of Oct. 24th 2003 }
\label{31024_diff}
\end{figure}

Do the differences in spectral index change in the course of the flare? Figure
\ref{31024_diff} displays the variations through the peak (at 02:48:30) and in the decay
phase. The difference between the footpoints' spectral indices does not vary
within the statistical error as given by the OSPEX routine. However, the
$\mathrm{\gamma_{cs} - \gamma_{fp}}$ increases from peak to decay. This is caused by a
considerable softening of the coronal source in this time interval (Fig. 4b).

\begin{figure}
\resizebox{\hsize}{!}{\includegraphics{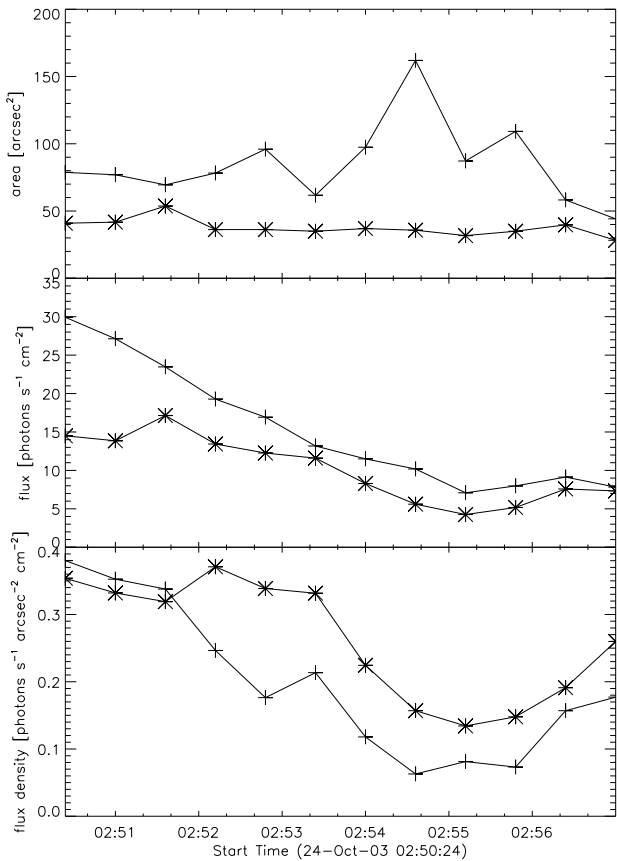}}
\caption {Time evolution of area of 50 \% contour in the 25 - 50 keV image, flux within contour in photons s$^{-1}$ cm$^{-2}$  and
  the intensity (flux/area) of footpoints of Oct. 24th 2003 event. \textit{Crosses} show footpoint no.1, \textit{asterisks} indicate footpoint no.2 as numbered in Fig. 1.}
\label{normflux}
\end{figure}

\subsubsection{Relation between intensity and area of footpoints in the Oct. 24th 2003 event}\label{fluxdensity}
 
We compared the total flux of non-thermal photons in the 25-50~keV range and
within the 50 \% contour for the two footpoints of the Oct. 24th 2003 event for the times where they were best observed. 

We determined the footpoint flux from cleaned images using detectors 3-8 and with a pixel size of 0.5 in the 25-50~keV range within the 50\% contour ($\mathrm{F_{50\%}}$) and define the intensity $\mathrm{f_{50\%}}$ as ratio of flux divided by the area ($\mathrm{A_{50\%}}$) of the contour,
$$I_{50\%}\ =\ {F_{50\%}\over A_{50\%}}\ \ ,$$ 
 where the area of the clean beam has been subtracted.
Figure~\ref{normflux}  shows the time evolution of $\mathrm{F_{50\%}}$, the area $\mathrm{A_{50\%}}$ and the intensity $\mathrm{I_{50\%}}$. 

The total flux F$_{50\%}$ correlates
with the full sun count flux. Further, it correlates with the fitted flux F$_{35}$ (comp. Fig.~\ref{SHS}), validating the applied methods. The softer footpoint, no. 1, is always
brighter than footpoint no. 2. and is larger in area. The harder footpoint (no. 2) has the higher intensity for most of the observed time.

\subsubsection{Thermal emission of footpoints}
The coronal source dominates at low energies in all selected events. Images at low energies often do not show strong emission at the position of the footpoints. Sometimes no fit to the thermal component of the footpoints was possible. As demonstrated in Fig.~\ref{samplespec}, the noise level at low energies (few keV up to 10~keV) reaches about a tenth of the emission of the coronal source. 

A rough overview on all times of all events yields the following statistics on the thermal emission of the footpoints:
\begin{itemize}
\item One footpoint with measurable thermal emission during more than 50~\% of the
  time (fp no. 1 of the Dec. 4th 2002 event). 
\item Two footpoints with no measurable thermal emission at all times (fp no.2 of the July 30th 2005 event and fp no. 2 of Oct. 24th 2003). 
\item All other footpoints show thermal emission in the spectrum for 20~\% of the time on the average. 
\end{itemize}
A formal fitting has been performed at the defined regions of interest for the undetected thermal footpoint emission in Fig.~\ref{samplespec}. The result corresponds to the uncertainty level and yields an upper limit for the thermal footpoint emission. The relations between the emission measure of the coronal source and the footpoints are 
$${\mathrm{EM_{fp1}\over EM_{cs}}}\ <\ 0.2$$ and
$${\mathrm{EM_{fp2}\over EM_{cs}}}\ <\ 0.1\ \ .$$ 
The fact that almost all of the footpoints studied here have a thermal
spectrum at some time and the well-known fact that 
flare observations at EUV wavelengths show thermal emission
from the footpoints suggest that there may well be thermal emission from footpoints all the time. In soft X-ray observations by \textit{Yohkoh/SXT}, such emission has been reported e.g. by McTiernan et al. (\cite{Mc93}) and Hudson et al. (\cite{Hu94}). However, RHESSI can only
observe it if the emission measure is at least 10 \% of the coronal source and at a temperature of several million Kelvin. 
Therefore, the thermal emission measured in full sun spectra is predominantly emission of the coronal source. Table~\ref{differences} gives the average temperatures as fitted to full sun spectra. They are representative for the temperature of the coronal source.


\section{Discussion}\label{discussion}
\subsection{Coronal source shows SHS-behavior}   
Although previously reported in the literature, the existence of a non-thermal
component in the coronal source is not trivial. As the thermal component is
strong and any non-thermal emission very soft, the latter is just an extension
at a much lower flux (see Fig. \ref{samplespec}c). We have tested the possibility of flux
pileup contributing to the range of energies where the non-thermal component
was fitted (section~\ref{pileup}). These tests show that the observed
HXR
 tail cannot be caused just by pileup but that there is significant
HXR source emission. We cannot exclude in all cases, however, that the extension cannot be fitted equally well with a second thermal component at a much higher temperature. In cases like Fig. \ref{samplespec}c, the fit with two thermal components has a higher $\chi$-square and is therefore less likely. We will thus continue to refer to the high-energy extension as non-thermal.

It is a remarkable result that for 5 out of 5 events, the time-evolution of the spectral index $\gamma$ of the coronal sources shows SHS-behavior. The event of July 13th 2005 is noteworthy, showing SHS-behavior in the coronal source, but not in the footpoints (Fig.\ref{SHS}d). The SHS-behavior of the coronal source would not be expected if SHS was just a transport effect such as Coulomb collisions or an electric field. 
Filtering of low energy electrons in the loop by collisions would not have an effect on the coronal source from where the particles may have originated. An induced electric field due to the return current $E = \eta j^{ret}$ (where $\eta$ is the electric resistivity), reflecting low energy particles from the loop back upwards would even lead to a softer spectrum in the coronal source, i.e. an anti-correlation between flux and spectral hardness. 

Although the notion of a pivot point was introduced by Grigis \& Benz (2004) as
a convenient and quantitative characterization of the SHS-behavior, Table \ref{differences}
suggests a possible physical significance of the pivot energy: $\mathrm{E^{piv}_{cs}}$
seems to increase with the temperature of the coronal source. The
significance needs to be confirmed by a larger sample. Another hint on a
possible physical relevance is the value of the pivot energy. In the flares in
which it could be determined, the pivot energy of the coronal source is at the energy (within the
error range) where the spectra of the thermal and non-thermal components
intersect (Fig. \ref{samplespec}c) or higher. Table \ref{differences} also shows that the pivot energy is an order of magnitude higher than the mean thermal energy. A deviation from a Maxwellian energy distribution or from isothermal homogeneity would be necessary to interpret the pivot energy as the starting point for the non-thermal acceleration. We do not consider it impossible, but highly speculative. 

\subsection{Soft-hard-soft behavior of the full sun}   
The soft-hard-soft (sometimes soft-hard-harder) behavior of solar flares has been extensively studied in full sun observations (see Introduction). As the coronal source usually dominates in the early phase of an event and remaines luminous longest, but at peak time the non-thermal part of full-sun spectra is dominated by the footpoints (e.g. Fig.~\ref{test}), the reported SHS results of full sun observations need to be tested for the possibility of spatial changes dominating temporal changes. 

\begin{figure}[!]
\resizebox{\hsize}{!}{\includegraphics{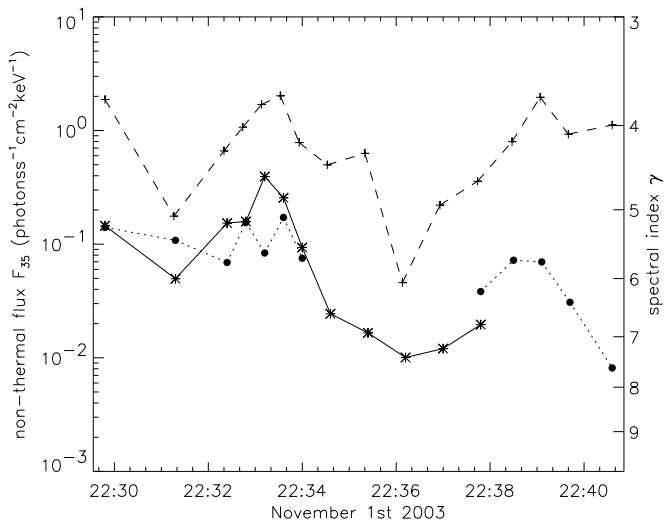}}
\caption {Non-thermal flux $\mathrm{F_{35}}$ fitted to the spectrum of the coronal source (dotted line) compared to summed flux of both footpoints (solid line) of the event of Nov. 1st 2003. The {\it dashed} line
gives the spectral index of the full sun spectra.}
\label{cfc}
\end{figure}

We have compared the time evolution of the non-thermal flux F$_{35}$ of the
coronal source and of the combined footpoints to investigate the influence of a change in predominance from the coronal source emission to footpoint emission and back to coronal emission  (coronal-footpoint-coronal) on the SHS feature. Figure~\ref{cfc} shows the only event where an indication of such an effect could be found. The footpoint emission is weaker than
the coronal source in the beginning, exceeds the coronal emission when the
spectral index is hardest and decreases below the level of coronal emission
afterwards. The (negative) spectral index of the full sun correlates with both the flux of footpoints and the flux of the coronal source. It continues to correlate with the coronal source when the footpoints vanish. 
We conclude that a coronal-footpoint-coronal effect may enhance
the SHS feature in full sun observations, but does not cause it. Therefore, the SHS-behavior must be a
property of the sources themselves.

\subsection{Differences in spectra}
\subsubsection{Difference between coronal source and footpoints}
Assuming an electron power-law distribution for the electron energy E of the form
\begin{equation} \label{electroneq}
F(E)=AE^{-\delta}
\end{equation}
producing
thin-target bremsstrahlung-emission in the coronal source, the
observed photon-spectrum at photon energies $\varepsilon$ is
\begin{equation} \label{thineq}
I_{thin}(\varepsilon)\sim \varepsilon^{-(\delta+1)} 
\end{equation}
with spectral index $\gamma_{thin}=\delta+1$.
Reaching the chromosphere, the accelerated electrons will be fully stopped, producing
thick-target bremsstrahlung with a photon-spectrum
\begin{equation} \label{thikeq}
I_{thick}(\varepsilon)\sim \varepsilon^{-(\delta-1)} 
\end{equation}
having a spectral index $\mathrm{\gamma_{thick}=\delta-1}$ (Tandberg-Hanssen \& Emslie \cite{Ta88}).
In such a simple scenario one would therefore expect a
difference in the photon spectral index $\gamma_{thin}-\gamma_{thick}=2$
between the coronal source and the footpoints.
Indeed we find always a difference between the spectral index of the coronal
source and the footpoints. In 2 events out of 5, the difference is considerably larger than 2. This result excludes a scenario in which the same electron beam first produces thin target emission in the corona, then thick target emission in the chromosphere. Different particle populations seem to be involved or a filter mechanism may operate causing low-energetic electrons to preferentially lose energy before they reach the chromosphere and therefore hardening the spectrum. Candidates for such a transport effect are collisions and the electric field of the return current.

The 3 events in which the difference is smaller than two include those two with small source separation. The similarities in the spectra can therefore be partly accounted for by a situation that is between the assumed ideal thick and thick targets. This may be the case when electrons substantially lose energy before reaching the chromosphere.\\

\subsubsection{Difference between footpoints}
The difference between the footpoint spectral indices is only significant in
one event out of five (Oct. 24th 2003). The larger and more luminous footpoint
is softer. The opposite is the case in the flare of July 23rd 2003, analyzed by Emslie et al. (2004).

\section{Conclusions}
A selection of five RHESSI events with three concurrent X-ray sources (coronal
source and two footpoints) has been studied regarding the spectral relations between the sources. All spectra can be fitted with a non-thermal component having a power-law photon distribution. Although no low-energy soft X-ray observations are available for comparison, we believe that they probably are examples for Masuda-type sources (Masuda \cite{Ma94}), but our looptop sources are generally softer than the one found by Masuda. Therefore, such events are easier to detect with RHESSI than they were with \textit{Yohkoh} and are more frequent than inferred previously.
In addition, all coronal sources and some of the footpoints at times show a thermal component. The major results are:

\begin{itemize}
\item All coronal sources evolve according to the same time evolution in spectral hardness. The higher the flux, the harder (smaller $\gamma$) the non-thermal component. This soft-hard-soft pattern correlates with the non-thermal flux without measurable delay. Transport effects such as collisions or an induced electric field cannot cause SHS in the coronal source. 
\item As the emission of the footpoints often dominates at 35~keV, it is not surprising that the pattern, previously reported for full sun observations, is also found in the footpoints of three out of 5 events. Imaging spectroscopy suggests that SHS is a feature of all sources, and thus possibly of the accelerator itself.
\item SHS in full sun observations cannot be explained by a change of the dominant source (softer coronal source present at all times plus a hard footpoint source with time-varying intensity). If SHS was caused by such an effect, neither source would display it individually.
\item The difference in spectral index between non-thermal coronal and footpoint emission is not 2, as would be the case if the difference was simply caused by thin and thick target bremsstrahlung, respectively. Smaller differences in $\gamma$ may be explained by a an intermediate situation between the two extremes. The plasma of the coronal source could act as thick target for low energetic electrons and as thin target for higher electron energies. The cases with $\gamma > 2$ require a filter effect in the propagation preferentially reducing the distribution at lower energies. Such a filter may be collisions or an electric field.  
\item The pivot energy, characterizing the SHS-behavior of the non-thermal emission, is at the energy where the distribution of the thermal and non-thermal components balance in half of the cases. In the other half, the pivot energy is higher than this point. 
\item The pivot energy at the footpoints is significantly lower in all cases
  ($16 -  23$ keV for the coronal source vs. $14 - 15$ keV for the footpoints). Such a difference suggests a filter acting during particle propagation to the footpoints, reducing lower energies more than higher energies.  
\item In one out of 5 events the two footpoints have significantly different spectral indices, $\Delta \gamma = 0.33\pm0.04$. The difference is constant during the event, although the spectral indices vary in time. Again, an energy filter during propagation seems to be at work, differing in one flare for the two legs of the loop. 
\item The photon flux at energies below about 15 keV is dominated by thermal emission. Most of this emission originates from the coronal source. If its temperature correlates with the pivot energy it may hint at a physical significance of the pivot energy for the acceleration process, but needs further investigation. 
\item As pointed out before from \textit{Yohkoh/SXT} observations, the thermal emission from the coronal source often significantly exceeds the thermal
  emission of the footpoints, which is detectable in some events and
  at some times. 
 \end{itemize}
This analysis has shown that the non-thermal X-ray emission in coronal sources cannot only be detected by RHESSI, but can also be studied in time. As the coronal source is directly related to flare energy release, this opens the possibility of further investigating the enigmatic acceleration process of electrons. The temporal and spectral relation of the coronal source to the footpoints suggests an intricate connection between corona and chromosphere. While a comprehensive interpretation of our results in terms of particle propagation and thermal conduction is beyond the scope of this study, the idea that flare energy release and particle acceleration are closely related to the coronal source is supported by our results.

\begin{acknowledgements}
RHESSI data analysis at ETH Z\"urich is supported by ETH grant TH-1/04-2 and the Swiss National Science Foundation (grant 20-105366). Much of this work relied on the RHESSI Experimental Data Center (HEDC), developed under ETH Z\"urich grant TH-W1/99-2. We thank S\"am Krucker, Paolo Grigis, Gordon Hurford and the unknown referee for helpful comments and discussions, and Andr\'e Csillaghy and Kim Tolbert for help with the software.
\end{acknowledgements}


\begin{thebibliography}{}
\bibitem[1997]{Al97} Alexander, D. \& Metcalf, T. R. 1997, \apj, 489, 442
\bibitem[2005]{Ba05} Battaglia, M., Grigis, P. C. \& Benz, A. O. 2005, \aap, 439, 737
\bibitem[1977]{Be77} Benz, A. O. 1977, \apj, 211, 270
\bibitem[1985]{Br85} Brown, J. C. \& Loran, J. M. 1985, MNRAS, 212, 245

\bibitem[1986]{Cl86} Cliver, E. W., Dennis, B. R., Kiplinger, A. L., et al. 1986, \apj, 305, 920
\bibitem[2003]{Em03} Emslie, G. A., Kontar, E. P., Krucker, S., Lin,
R. P. 2003 \apj, 595, L107
\bibitem[2002]{Fl02} Fletcher, L. \& Hudson, H. S. 2002, \solphys, 210, 307
\bibitem[2004]{Fl04} Fletcher, L., Pollock, J. A. \& Potts, H. E. 2004,
  \solphys, 222, 279
\bibitem[1971]{Fr71} Frost, K. J. \& Dennis, B. R. 1971, \apj, 165, 655
\bibitem[2000]{Ga00} Gallagher, P. T., Williams, D. R., Phillips, K. J. H., et
  al. 2000, \solphys, 195, 367
\bibitem[2004]{Gr04} Grigis, P. C. \& Benz, A. O. 2004, \aap, 426, 1093
\bibitem[2005]{Gr05} Grigis, P. C. \& Benz, A. O. 2005, \aap, 434, 1173
\bibitem[1981]{Ho81} Hoyng, P., Duijveman, A., Machado, M.E., et al. 1981
  \apj, 246, L155
\bibitem[2002]{Hud02} Hudson, H. S. \& F\'arn\'ik, F. 2002, ESA SP-505: Solar Variability: From Core to Outer Frontiers, 261
\bibitem[1994] {Hu94} Hudson, H. S., Strong, K. T., Dennis, B. R., et al. 1994, \apj, 422, L25
\bibitem[2002] {Hu02} Hurford, G. J., Schmahl, E. J., Schwartz, R. A., et
  al. 2002, \solphys, 210, 61
\bibitem[1998] {Ga98} Gan, W. Q. 1998, \apss, 260, 515
\bibitem[1970] {Ka70} Kane, S. R. \& Anderson, K. A. 1970, \apj, 162, 1003
\bibitem[1995]{Ki95} Kiplinger, A. L. 1995, \apj, 453, 973
\bibitem[1986]{Li86} Lin, R. P. \& Schwartz, R. A. 1986, \apj, 312, 462
\bibitem[2002] {Li02} Lin, R. P., Dennis, B. R., Hurford, G. J., et al. 2002,
  \solphys, 210, 3
\bibitem[1994]{Ma94} Masuda, S. , Kosugi, T., Hara, H., et al. 1994 \nat,
  371, 495
\bibitem[1993]{Mc93} McTiernan, J. M., Kane, S. R., Loran, J. M., et al. 1993, \apj, 416, L91
\bibitem[1969]{Pa69} Parks, G. K. \& Winckler, J. R. 1969, \apj, 155, 117
\bibitem[2002]{Pe02} Petrosian, V., Donaghy, T. Q. \& McTiernan, J. M. 2002, \apj, 569, 459
\bibitem[2002]{Sa02} Saint-Hilaire, P., von Praun, C., Stolte, E., et
  al. 2002, \solphys, 210, 143
\bibitem[1996]{Sc96} Schwartz, R. A. 1996, Compton Gamma Ray Observatory Phase 4 Gues Investigator Program: Solar Flare Hard X-ray Spectroscopy, Technical Report, NASA Goddard Space Flight Center
\bibitem[2002]{Sm02} Smith, D. M., Lin, R. P., Turin, P., et al. 2002
  \solphys, 210, 33 
\bibitem[1988]{Ta88} Tandberg-Hanssen, E. \& Emslie, A. G. 1988, The
  Physics of Solar Flares, Cambridge Astrophysics Series
\end{thebibliography}
\end{document}